\documentclass[12pt,a4paper]{article}

\usepackage[british]{babel}
\usepackage[a4paper,top=2cm,bottom=2cm,left=2.5cm,right=2.5cm,marginparwidth=1.75cm]{geometry}

\usepackage{amsmath}
\usepackage{graphicx}
\usepackage[colorlinks=true, allcolors=blue]{hyperref}
\usepackage{hyperref}
\usepackage[title]{appendix}
\usepackage{mathrsfs}
\usepackage{amsfonts}
\usepackage{booktabs} 
\usepackage{caption}  
\usepackage{threeparttable} 
\usepackage{algorithm}
\usepackage{algorithmicx}
\usepackage{algpseudocode}
\usepackage{listings}
\usepackage{enumitem}
\usepackage{chngcntr}
\usepackage{booktabs}
\usepackage{lipsum}
\usepackage{subcaption}
\usepackage{authblk}
\usepackage[T1]{fontenc}    
\usepackage{csquotes}       
\usepackage{diagbox}
\usepackage{cite}

\usepackage{setspace}
\usepackage{titlesec}
\titleformat{\section} 
  {\normalfont\Large\bfseries}{\thesection.}{1em}{}
  
\usepackage{lineno} 

\rightlinenumbers 

\usepackage{float}   
\usepackage{caption} 


\title{Deep Learning in Long-Short Stock Portfolio Allocation: An Empirical Study}

\author[1,*]{Junjie Guo}
\affil{\small Rutgers University, New Jersey, USA}
\affil[*]{Corresponding author: \texttt{jg1806@scarletmail.rutgers.edu}}

\date{}  

\begin{document}
\maketitle

\begin{abstract}
This paper provides an empirical study explores the application of deep learning algorithms—Multilayer Perceptron (MLP), Convolutional Neural Networks (CNN), Long Short-Term Memory (LSTM), and Transformer—in constructing long-short stock portfolios. Two datasets comprising randomly selected stocks from the S\&P 500 and NASDAQ indices, each spanning a decade of daily data, are utilized. The models predict daily stock returns based on historical features such as past returns, Relative Strength Index (RSI), trading volume, and volatility. Portfolios are dynamically adjusted by longing stocks with positive predicted returns and shorting those with negative predictions, with equal asset weights. Performance is evaluated over a two-year testing period, focusing on return, Sharpe ratio, and maximum drawdown metrics. The results demonstrate the efficacy of deep learning models in enhancing long-short stock portfolio performance.
\end{abstract}

\section{Introduction}
The rapid advancement of deep learning techniques has opened new avenues in financial modeling and asset management\cite{Cheng2024Advanced,Xu2024Advancing}. Traditional quantitative methods often rely on linear assumptions and may not capture the complex, nonlinear patterns inherent in financial markets. Deep learning models, with their ability to learn intricate patterns from large datasets, present an opportunity to improve prediction accuracy and, consequently, enhance investment strategies\cite{guo2020news}.

Stock portfolio allocation is a critical aspect of investment management, aiming to balance risk and return effectively. The long-short portfolio strategy, in particular, allows investors to capitalize on both overvalued and undervalued securities\cite{ding2015deep,ritter2017machine}. By integrating deep learning models into this strategy, it becomes possible to refine return predictions and optimize portfolio performance.

This study conducts an empirical analysis of deep learning applications in long-short stock portfolio allocation. Focusing on four prominent deep learning architectures -Multilayer Perceptron (MLP), Convolutional Neural Networks (CNN), Long Short-Term Memory (LSTM), and Transformers—we train these models on historical stock data from two major indices: the S\&P 500 and NASDAQ. The selected stocks provide a decade of daily data, enabling the models to predict daily returns using features such as past returns, Relative Strength Index (RSI), trading volume, and volatility. The predicted returns inform the construction of dynamically adjusted portfolios.

By evaluating the performance of these deep learning models in predicting stock returns and their impact on portfolio allocation, this research contributes to the growing body of literature on machine learning applications in finance. The study offers practical insights for portfolio managers seeking to leverage deep learning techniques, focusing on metrics such as return, Sharpe ratio, and maximum drawdown over a two-year testing period.

\section{Related Works}\label{sec2}
The intersection of deep learning and financial markets has attracted significant attention in recent years. Deep learning models have been applied to various aspects of finance, including price prediction, risk assessment, and portfolio optimization.

Previous studies have employed deep learning models for stock price and return predictions. LSTM networks have been utilized to predict stock movements in the S\&P 500 index, demonstrating that LSTM models outperform traditional methods like logistic regression \cite{fischer2018deep}. Similarly, the effectiveness of wavelet transforms, stacked autoencoders, and LSTM networks in stock price prediction was compared, finding that deep learning models offer superior performance due to their ability to capture temporal dependencies and nonlinear relationships \cite{bao2017deep}.

Deep learning has also been explored in the context of portfolio optimization. Researchers discussed how deep learning models could be integrated into asset allocation strategies, highlighting the potential for these models to learn complex patterns in financial data \cite{heaton2017deep}. Deep reinforcement learning for portfolio management was introduced, proposing a framework that allows for continuous learning and adaptation to market changes \cite{jiang2017deep}.

The choice of input features significantly impacts the performance of deep learning models in finance. Studies emphasized the importance of technical indicators, such as RSI and moving averages, as inputs to neural networks for stock trend prediction \cite{patel2015predicting,chong2017deep}. These features help models capture market momentum and investor behavior patterns.

Research comparing different deep learning architectures in financial applications is relatively scarce. However, a comprehensive review of deep learning in stock market prediction noted that while CNNs are effective in extracting spatial features, LSTMs are better suited for sequential data common in financial time series \cite{sezer2020financial}.

The existing literature underscores the potential of deep learning models in improving financial predictions and portfolio management. However, there is a gap in empirical studies that compare multiple deep learning architectures within the context of long-short portfolio allocation. This study aims to fill that gap by evaluating the performance of MLP, CNN, LSTM, and Transformer models in predicting stock returns and enhancing portfolio outcomes.

\section{Preliminary}\label{sec3}
In this section, we present the theoretical foundations pertinent to our study, including deep learning approaches for stock return prediction, the construction of long-short portfolios, and key performance metrics such as the Sharpe ratio, portfolio return, and maximum drawdown.

\subsection{Deep Learning-Based Stock Return Prediction}
Deep learning models are powerful tools capable of capturing complex, nonlinear relationships in financial time series data. In the context of stock return prediction, these models learn patterns from historical data to forecast future returns.
In supervised deep learning, the objective is to learn a function \( f_{\theta} \) that maps input data \( \mathbf{x} \) to output targets \( \mathbf{y} \), based on a set of labeled training examples. The function \( f_{\theta} \) is parameterized by a set of parameters \( \theta \), which are optimized during training\cite{hastie2009elements}.

Given a training dataset \( \mathcal{D} = \{ (\mathbf{x}^{(i)}, \mathbf{y}^{(i)}) \}_{i=1}^{N} \), where \( \mathbf{x}^{(i)} \in \mathbb{R}^{n} \) is the input feature vector, \( \mathbf{y}^{(i)} \in \mathbb{R}^{m} \) is the corresponding target output.A deep learning model can be described as a mapping  \( f_{\theta}: \mathbb{R}^{n} \rightarrow \mathbb{R}^{m} \),parameterized by \( \theta \). In the context of stock return prediction $m$ equals one which is a scalar.     

In this formulation, let \( \mathbf{X}_t \) represent the feature vector at time \( t \), the objective is to predict the next day's return \( r_{t+1} \):
\begin{equation}
    \hat{r}_{t+1} = f_{\theta}(\mathbf{X}_t)\
\end{equation}

\subsection{Long-Short Stock Portfolio Construction}
A long-short portfolio strategy involves taking long positions in stocks expected to increase in value and short positions in stocks expected to decrease. Based on the predicted returns \( \hat{r}_{t+1} \), stocks are categorized as two groups.The long positions (\( \mathcal{L} \)) include stocks with \( \hat{r}_{t+1} > 0 \).The short positions (\( \mathcal{S} \))include stocks with \( \hat{r}_{t+1} < 0 \).The return of the long-short portfolio on day \( t+1 \) can be calculated as:
\begin{equation}
    \ R_{t+1} = \sum_{i \in \mathcal{L}} w_i \, r_{i,t+1} - \sum_{i \in \mathcal{S}} w_i \, r_{i,t+1}\
\end{equation}
where \( r_{i,t+1} \) is the actual return of stock \( i \) at time \( t+1 \).

\subsection{Sharpe Ratio}

The Sharpe ratio measures the performance of an investment by adjusting for its risk. It is defined as\cite{Sharpe1966}:
\begin{equation}
    \text{Sharpe Ratio} = \frac{E[R_p] - R_f}{\sigma_p}\
\end{equation}

where \( E[R_p] \) is the expected portfolio return, \( R_f \) is the risk-free rate, \( \sigma_p \) is the standard deviation of portfolio returns.

In practice, \( E[R_p] \) and \( \sigma_p \) are estimated from historical data:
\begin{equation}
    \ E[R_p] = \frac{1}{T} \sum_{t=1}^{T} R_{t} \
\end{equation}

\begin{equation}
    \sigma_p = \sqrt{ \frac{1}{T-1} \sum_{t=1}^{T} \left( R_{t} - E[R_p] \right)^2 } \
\end{equation}

where \( T \) is the number of periods.

\subsection{Return of Long-Short Portfolio}

The cumulative return of the portfolio over the testing period is calculated as:
\begin{equation}
    \text{Cumulative Return} = \prod_{t=1}^{T} (1 + R_{t}) - 1 \
\end{equation}

where \( R_{t} \) is the portfolio return at time \( t \).

The average (arithmetic) return is:
\begin{equation}
    \overline{R_p} = \frac{1}{T} \sum_{t=1}^{T} R_{t}\
\end{equation}

\subsection{Maximum Drawdown}

Maximum drawdown (MDD) quantifies the largest loss from a peak to a trough before a new peak is attained. It is a measure of downside risk over a specified time period\cite{DeRosa1999,AiyagariWhite2000}.

Let \( V_t \) be the cumulative portfolio value at time \( t \). The drawdown at time \( t \) is:
\begin{equation}
    \text{Drawdown}_t = \frac{V_t - \max\limits_{s \leq t} V_s}{\max\limits_{s \leq t} V_s}\
\end{equation}

The maximum drawdown over the period is:
\begin{equation}
    \text{Maximum Drawdown} = \min\limits_{t \in [1, T]} \left( \text{Drawdown}_t \right)\
\end{equation}

Maximum drawdown is expressed as a negative percentage, indicating the maximum observed loss from peak to trough. We take the absolute value of this maximum drawdown to get a positive outcome. The larger the value, the worse the performance of the portfolio over the backtesting period.

\section{Methods}\label{sec4}
\subsection{Data Collection}

We selected two groups of stocks: one from the S\&P 500 index and another from the NASDAQ index. For each group, 10 stocks were randomly chosen, ensuring each had 10 years of continuous daily stock price data from October 1, 2014, to October 25, 2024. The data was sourced from Yahoo Finance. We split the 10 years data into training and test set. Training set contains data from October 1, 2014, to October 24, 2022. Test set contains data from October 25, 2022, to October 25, 2022.
Table 1 illustrate the tickers of the selected stocks.

\begin{table}[h!]
    \centering
    \begin{tabular}{|c|c|}
        \hline
        \textbf{S\&P 500 Tickers} & \textbf{NASDAQ Tickers} \\
        \hline
        AAPL & FISV \\
        PFE  & MRNA \\
        ADBE & BKNG \\
        MSFT & INTC \\
        PG   & NFLX \\
        JPM  & INTU \\
        HD   & AMAT \\
        AMZN & AVGO \\
        KO   & CMCSA \\
        NFLX & PEP \\
        \hline
    \end{tabular}
    \caption{Selected Tickers from S\&P 500 and NASDAQ}
    \label{tab:tickers}
\end{table}

Figure 1 shows the original price data obtained. Within the observed period, certain stocks exhibited substantial price appreciation, reflecting notable upward momentum and possible increased investor interest. In contrast, other stocks demonstrated price stability, with minimal fluctuations, indicating relatively balanced market demand and supply dynamics. 

\begin{figure}[!ht]
    \begin{subfigure}{0.5\textwidth}
        \includegraphics[width=\linewidth]{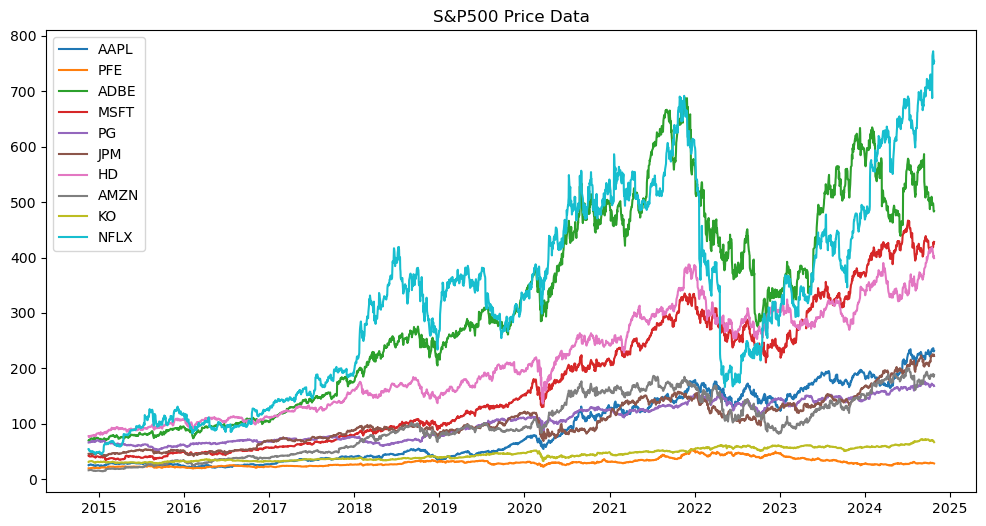}
        \caption{}
    \end{subfigure}
    \hfill
    \begin{subfigure}{0.5\textwidth}
        \includegraphics[width=\linewidth]{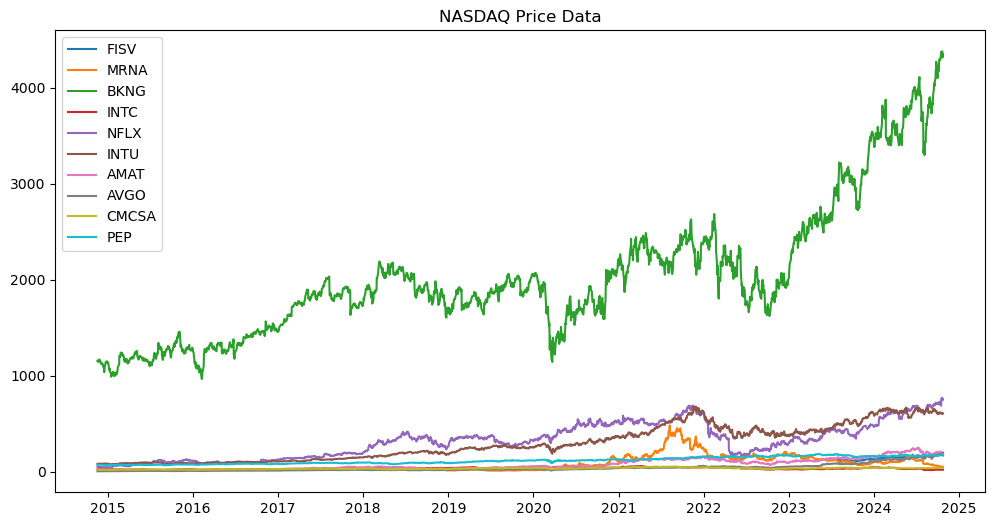}
        \caption{}
    \end{subfigure}
    \hfill
    \caption{Price data of S\&P 500 and NASDAQ stocks}
    \label{fig:multi_figs}
\end{figure}

\subsection{Feature Engineering}
In this paper, we use four features as input of the model to predict the return for each stock. The four features are Return, RSI, Volume and volatility. 

In this paper, we use daily return as one feature. The daily return means the change ratio of today's price compared to previous price. At any given day \( t \), the return can be calculated as in (12), where \( P_t \) is the closing price on day \( t \). Figure 2 shows the return data of selected stocks.
\begin{figure}[!ht]
    \begin{subfigure}{0.5\textwidth}
        \includegraphics[width=\linewidth]{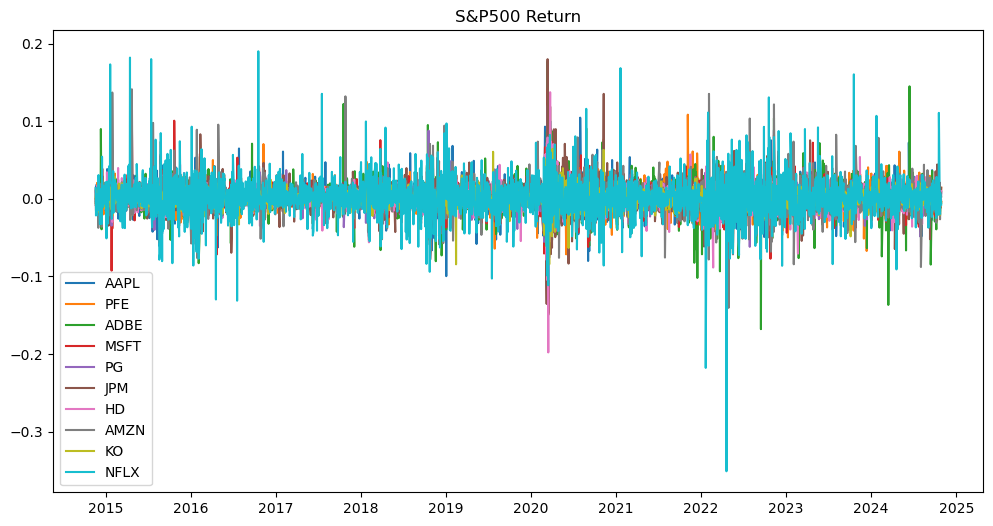}
        \caption{}
    \end{subfigure}
    \hfill
    \begin{subfigure}{0.5\textwidth}
        \includegraphics[width=\linewidth]{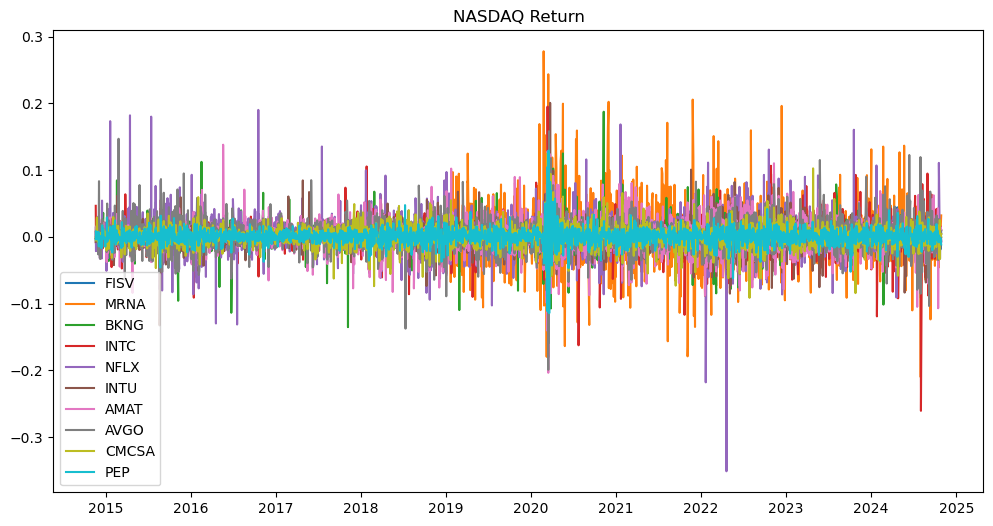}
        \caption{}
    \end{subfigure}
    \hfill
    \caption{Return data of S\&P 500 and NASDAQ stocks}
    \label{fig:multi_figs}
\end{figure}

\begin{equation}
R_t = \frac{P_t - P_{t-1}}{P_{t-1}}
\end{equation}

As a momentum oscillator measuring the speed and change of price movements, RSI values range between 0 and 100, providing insights into overbought or oversold conditions of a security. An RSI value above 70 typically indicates an overbought scenario, suggesting that the asset may be overvalued and poised for a price correction. Conversely, an RSI below 30 signals an oversold condition, implying potential undervaluation and an upcoming price increase. By identifying these extremities in price momentum, RSI enables investors to anticipate reversals in stock price trends, thereby serving as a valuable feature in predicting stock returns. The RSI over 14 days can be calculated as in (13). Figure 3 shows the calculated RSI of selected stocks.

\begin{equation}
\text{RSI}_t = 100 - \left( \frac{100}{1 + \frac{\frac{1}{14} \sum_{i=1}^{14} \max(R_{t - i}, 0)}{\frac{1}{14} \sum_{i=1}^{14} \max(-R_{t - i}, 0)}} \right)
\end{equation}

\begin{figure}[!ht]
    \begin{subfigure}{0.5\textwidth}
        \includegraphics[width=\linewidth]{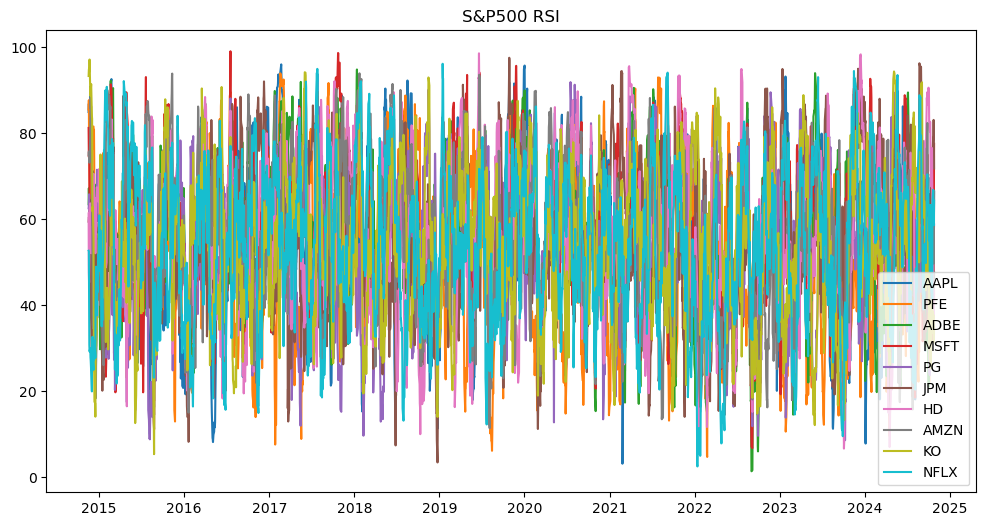}
        \caption{}
    \end{subfigure}
    \hfill
    \begin{subfigure}{0.5\textwidth}
        \includegraphics[width=\linewidth]{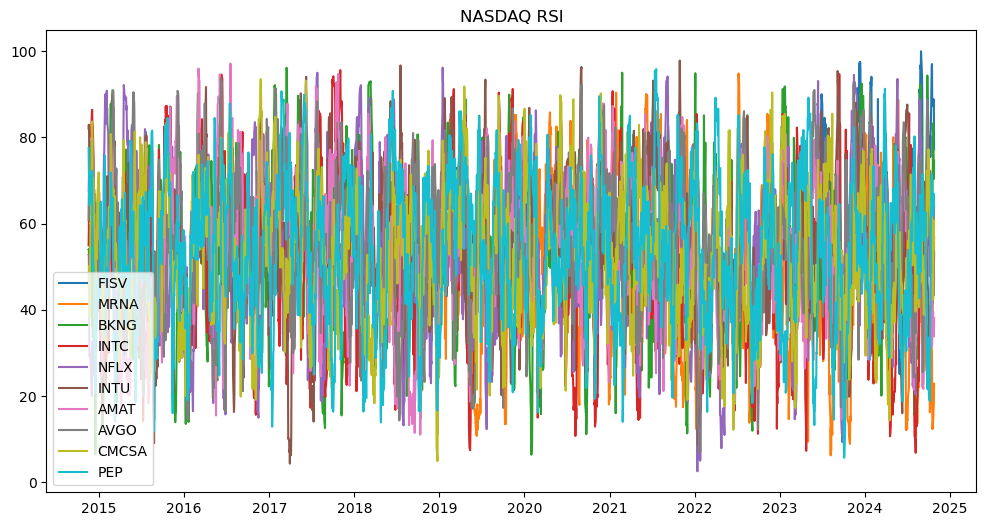}
        \caption{}
    \end{subfigure}
    \hfill
    \caption{RSI data of S\&P 500 and NASDAQ stocks}
    \label{fig:multi_figs}
\end{figure}

Trading volume is another pivotal factor in predicting stock returns, as it reflects the intensity of market participation and investor interest in a particular security. High trading volumes are often associated with significant price movements and can confirm the strength of a prevailing trend. For instance, a price increase accompanied by high volume suggests strong buying interest and the sustainability of the upward movement. Conversely, low volume during price changes may indicate weak momentum and the possibility of a trend reversal. Figure 4 shows the volume data of selected stocks.
\begin{figure}[!ht]
    \begin{subfigure}{0.5\textwidth}
        \includegraphics[width=\linewidth]{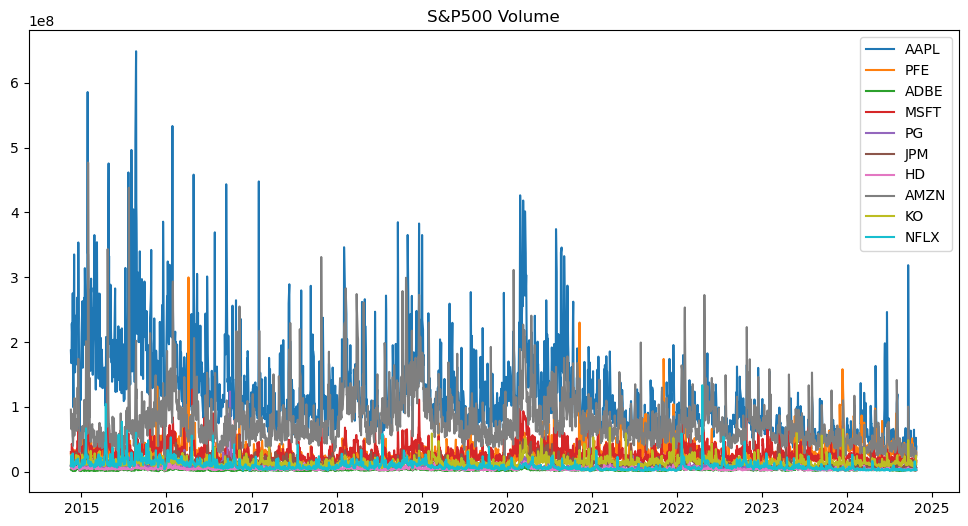}
        \caption{}
    \end{subfigure}
    \hfill
    \begin{subfigure}{0.5\textwidth}
        \includegraphics[width=\linewidth]{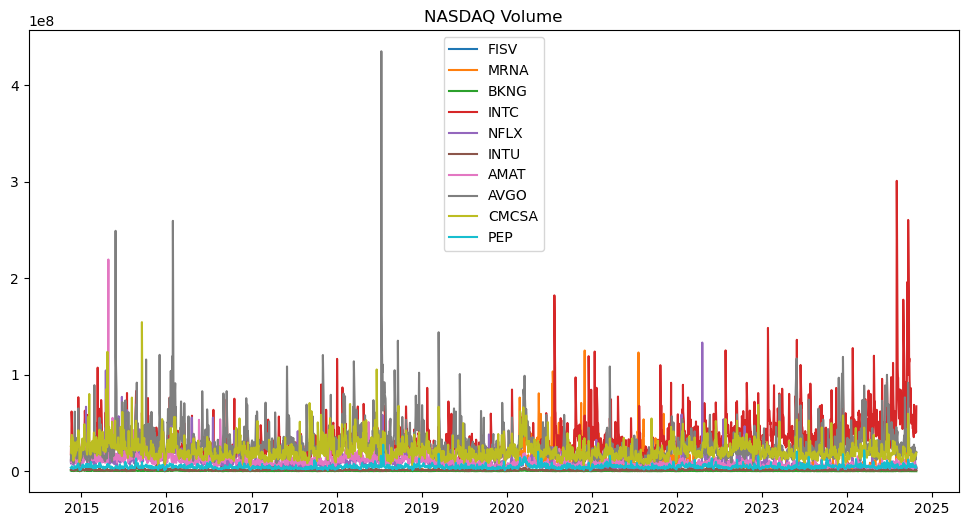}
        \caption{}
    \end{subfigure}
    \hfill
    \caption{Volume data of S\&P 500 and NASDAQ stocks}
    \label{fig:multi_figs}
\end{figure}

Volatility of return is essential in understanding the risk and uncertainty inherent in a stock's price fluctuations over time. It quantifies the degree of variation in returns, with higher volatility indicating larger price swings and a higher risk profile. Stocks with high volatility present both opportunities for significant gains and risks of substantial losses. Incorporating volatility into predictive models allows investors to evaluate the risk-adjusted potential of a stock, aiding in the development of investment strategies that align with their risk tolerance. At any time \( t \), the volatility of return over the past 5 days can be calculated as in (16), where \( \bar{R} \) is the mean return over the 5-day period.Figure 5 shows the volatility data of selected stocks.

\begin{equation}
\sigma_t = \sqrt{\frac{1}{4} \sum_{i=1}^{5} \left( R_{t - i} - \bar{R} \right)^2}
\end{equation}

\begin{figure}[!ht]
    \begin{subfigure}{0.5\textwidth}
        \includegraphics[width=\linewidth]{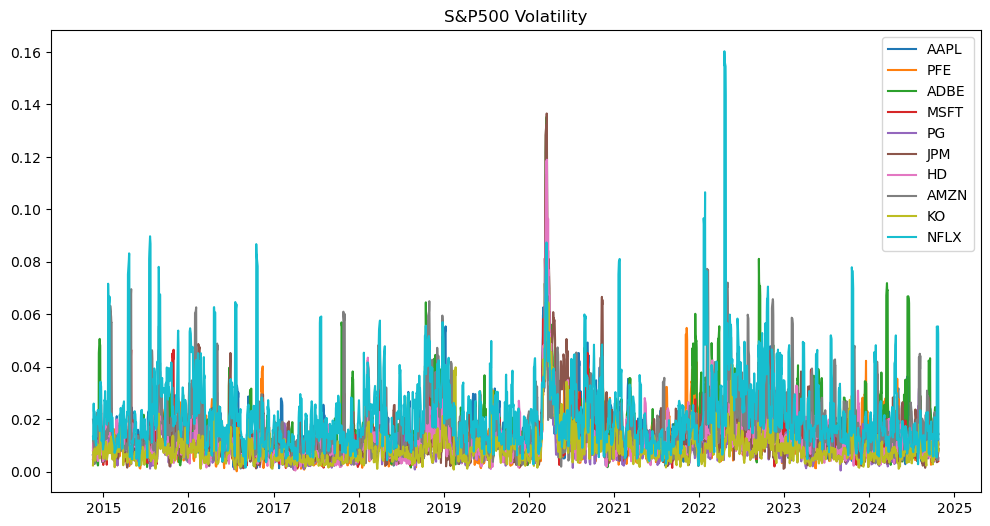}
        \caption{}
    \end{subfigure}
    \hfill
    \begin{subfigure}{0.5\textwidth}
        \includegraphics[width=\linewidth]{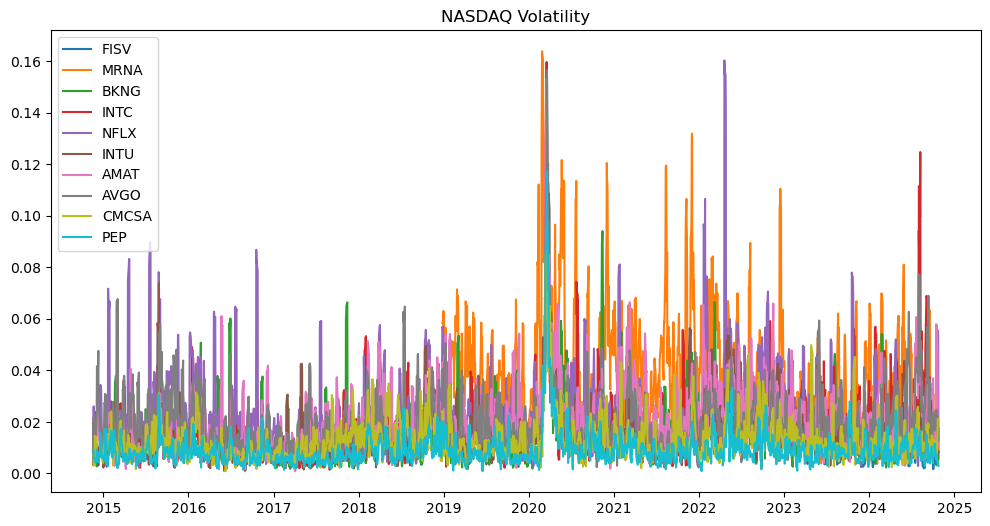}
        \caption{}
    \end{subfigure}
    \hfill
    \caption{Volatility data of S\&P 500 and NASDAQ stocks}
    \label{fig:multi_figs}
\end{figure}

\subsection{Deep Learning Models}

In this empirical study, we employ several deep learning architectures to predict daily stock returns based on historical features. The models utilized include the Multilayer Perceptron (MLP), Convolutional Neural Network (CNN), Long Short-Term Memory (LSTM) networks, and Transformer models. 

The Multilayer Perceptron is a class of feedforward artificial neural networks consisting of multiple layers of nodes in a directed graph, with each layer fully connected to the next one. The MLP is capable of modeling complex non-linear relationships between inputs and outputs, making it suitable for stock return prediction.The forward propagation in an MLP can be formulated as:

\begin{equation}
    \begin{aligned}
        \mathbf{a}^{(0)} &= \mathbf{x} \\
        \mathbf{z}^{(l)} &= \mathbf{W}^{(l)} \mathbf{a}^{(l-1)} + \mathbf{b}^{(l)} \\
        \mathbf{a}^{(l)} &= \phi(\mathbf{z}^{(l)})
    \end{aligned}
\end{equation}

where $\mathbf{x} \in \mathbb{R}^n$ is the input feature vector; $\mathbf{W}^{(l)} \in \mathbb{R}^{m_l \times m_{l-1}}$ and $\mathbf{b}^{(l)} \in \mathbb{R}^{m_l}$ are the weight matrix and bias vector for layer $l$; $\mathbf{a}^{(l)}$ is the activation vector of layer $l$; and $\phi(\cdot)$ is the activation function (e.g., ReLU, sigmoid).

Convolutional Neural Networks are designed to process data with grid-like topology, such as time series data. In this study, a one-dimensional CNN is utilized to capture local temporal patterns in the stock features. The convolution operation in a CNN is defined as\cite{krizhevsky2012imagenet}:

\begin{equation}
    \mathbf{z}_i^{(l)} = (\mathbf{w}^{(l)} * \mathbf{a}^{(l-1)})_i + b^{(l)} = \sum_{k=0}^{K-1} \mathbf{w}_k^{(l)} \mathbf{a}_{i+k}^{(l-1)} + b^{(l)} \
\end{equation}

where $\mathbf{w}^{(l)} \in \mathbb{R}^K$ is the convolution kernel of size $K$; $\mathbf{a}^{(l-1)}$ is the input activation from the previous layer; $b^{(l)}$ is the bias term; and $\mathbf{z}_i^{(l)}$ is the output at position $i$ in layer $l$.

LSTM networks are a type of recurrent neural network (RNN) capable of learning long-term dependencies in sequential data, making them effective for time series forecasting.The computations within an LSTM cell are as follows\cite{hochreiter1997long}:
\begin{equation}
    \begin{aligned}
    \mathbf{f}_t &= \sigma(\mathbf{W}_f \mathbf{x}_t + \mathbf{U}_f \mathbf{h}_{t-1} + \mathbf{b}_f) \\
    \mathbf{i}_t &= \sigma(\mathbf{W}_i \mathbf{x}_t + \mathbf{U}_i \mathbf{h}_{t-1} + \mathbf{b}_i) \\
    \mathbf{o}_t &= \sigma(\mathbf{W}_o \mathbf{x}_t + \mathbf{U}_o \mathbf{h}_{t-1} + \mathbf{b}_o) \\
    \mathbf{c}_t &= \mathbf{f}_t \odot \mathbf{c}_{t-1} + \mathbf{i}_t \odot \tanh(\mathbf{W}_c \mathbf{x}_t + \mathbf{U}_c \mathbf{h}_{t-1} + \mathbf{b}_c) \\
    \mathbf{h}_t &= \mathbf{o}_t \odot \tanh(\mathbf{c}_t)
    \end{aligned}
\end{equation}

where $\mathbf{x}_t$ is the input at time $t$; $\mathbf{h}_t$ is the hidden state at time $t$; $\mathbf{c}_t$ is the cell state at time $t$; $\sigma(\cdot)$ is the sigmoid activation function; $\odot$ denotes element-wise multiplication; and $\mathbf{W}_*$, $\mathbf{U}_*$, $\mathbf{b}_*$ are weight matrices and bias vectors.

The Transformer model, introduced by \cite{vaswani2017attention}, relies entirely on attention mechanisms to model dependencies in sequential data without using recurrence or convolution.The self-attention mechanism is defined as:
\begin{equation}
    \text{Attention}(Q, K, V) = \text{softmax}\left(\frac{Q K^\top}{\sqrt{d_k}}\right) V \
\end{equation}

where $Q = \mathbf{X} \mathbf{W}_Q$, $K = \mathbf{X} \mathbf{W}_K$, $V = \mathbf{X} \mathbf{W}_V$; $\mathbf{X}$ is the input sequence; $\mathbf{W}_Q$, $\mathbf{W}_K$, $\mathbf{W}_V$ are the weight matrices for queries, keys, and values; and $d_k$ is the dimension of the key vectors.

\section{Experiment and result}

The two groups of stocks are trained using four models on the training set.Those trained models then utilized on test set to backtest the performance of long-short portfolio strategy. At each day of test set, we predict the next day's return. We long those stocks with positive return and short those stocks with negative return. Assume we fully invest the portfolio. The weights of investing in each stock are equally assigned. The weight vector can be defined as:
\begin{equation}
    \mathbf{W}_p = \frac{1}{N}\mathbf{I}
\end{equation}

Where $\mathbf{I}$ has all elements equal to 1 and same dimension as $\mathbf{W}_p$.

Portfolios contain the same stocks at each day, but are rebalanced based on new predictions and weights. We assume transaction costs are negligible. After backtesting, we can get returns for the two groups of stocks for each model. Figure 6 shows the returns of stocks in the S\&P 500 group over the backtesting period. Figure 7 shows the returns of stocks in the NASDAQ group over the backtesting period.

\begin{figure}[!ht]
    \begin{subfigure}{0.5\textwidth}
        \includegraphics[width=\linewidth]{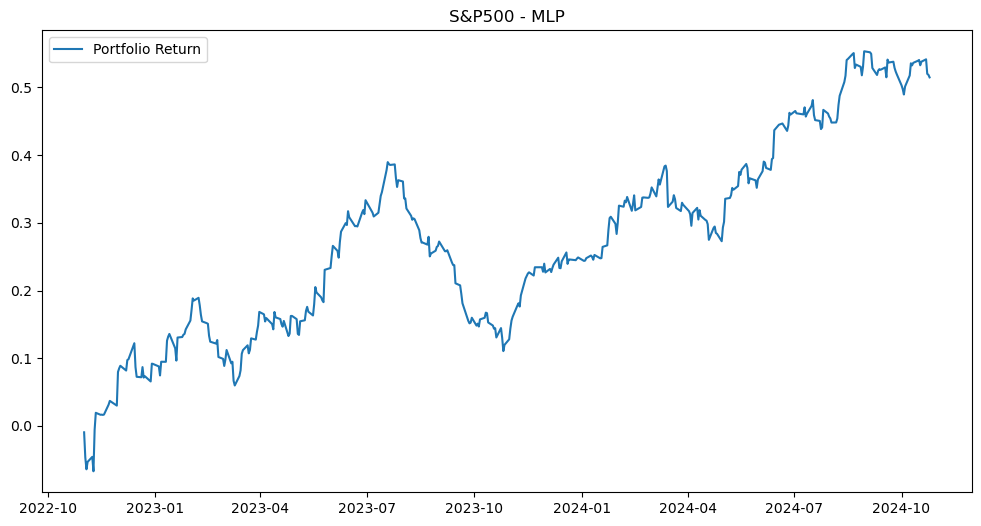}
        \caption{}
    \end{subfigure}
    \hfill
    \begin{subfigure}{0.5\textwidth}
        \includegraphics[width=\linewidth]{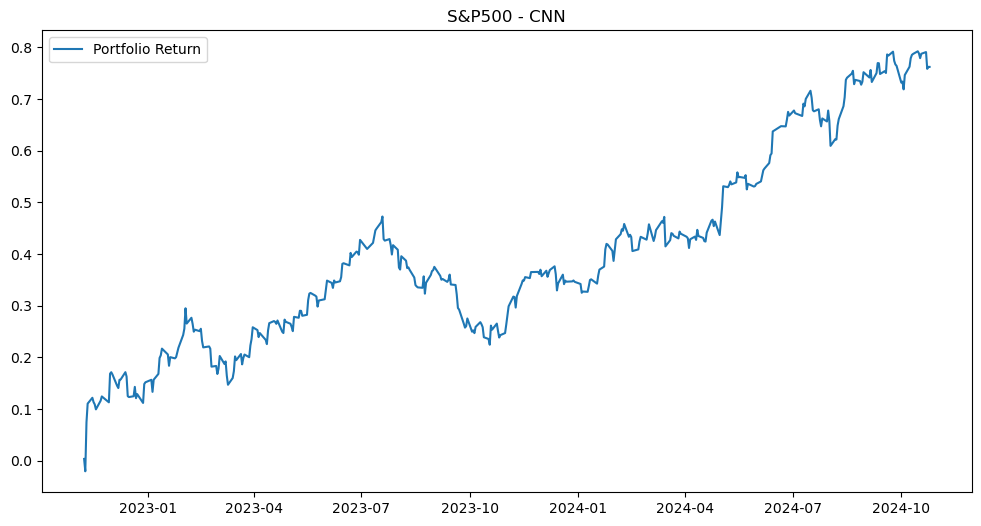}
        \caption{}
    \end{subfigure}
    \hfill
    \begin{subfigure}{0.5\textwidth}
        \includegraphics[width=\linewidth]{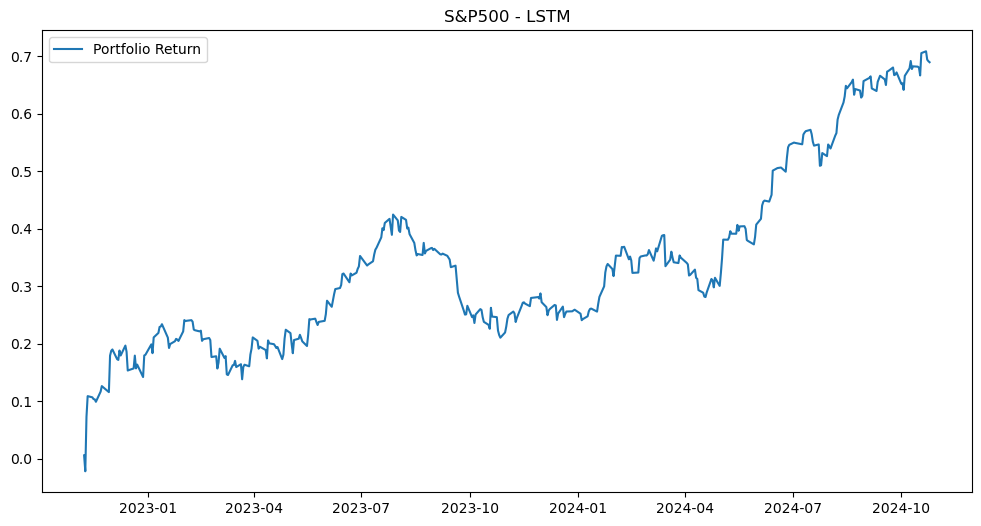}
        \caption{}
    \end{subfigure}
    \hfill
    \begin{subfigure}{0.5\textwidth}
        \includegraphics[width=\linewidth]{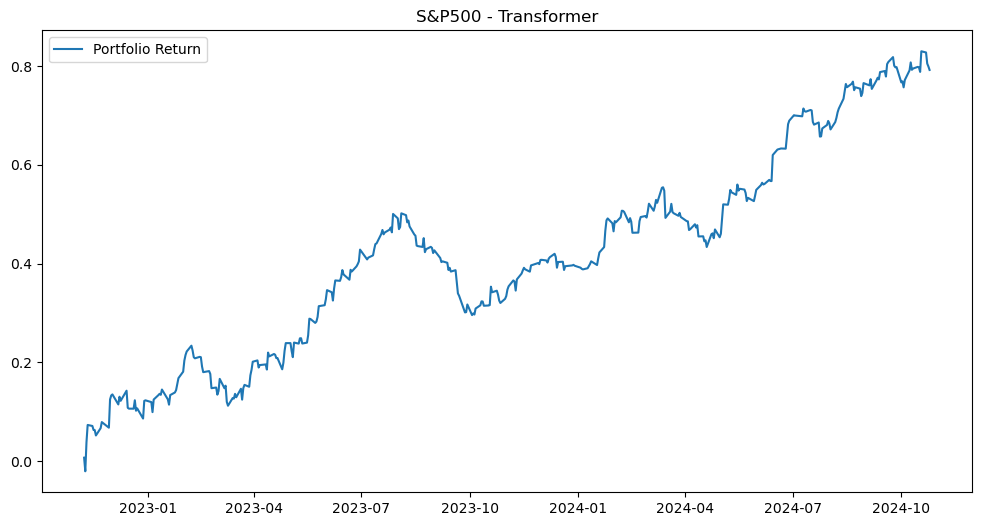}
        \caption{}
    \end{subfigure}
    \hfill
    \caption{Total return of S\&P 500 stocks}
    \label{fig:multi_figs}
\end{figure}

\begin{figure}[!ht]
    \begin{subfigure}{0.5\textwidth}
        \includegraphics[width=\linewidth]{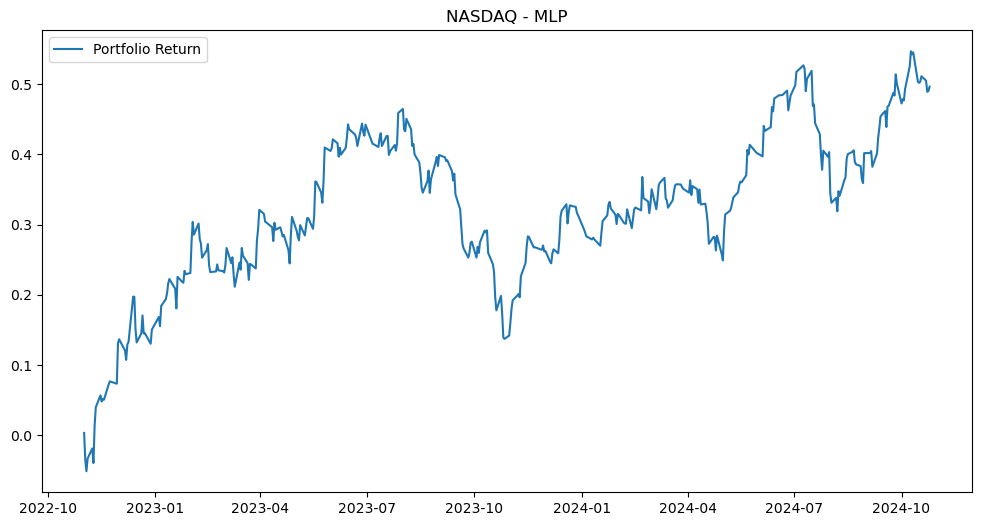}
        \caption{}
    \end{subfigure}
    \hfill
    \begin{subfigure}{0.5\textwidth}
        \includegraphics[width=\linewidth]{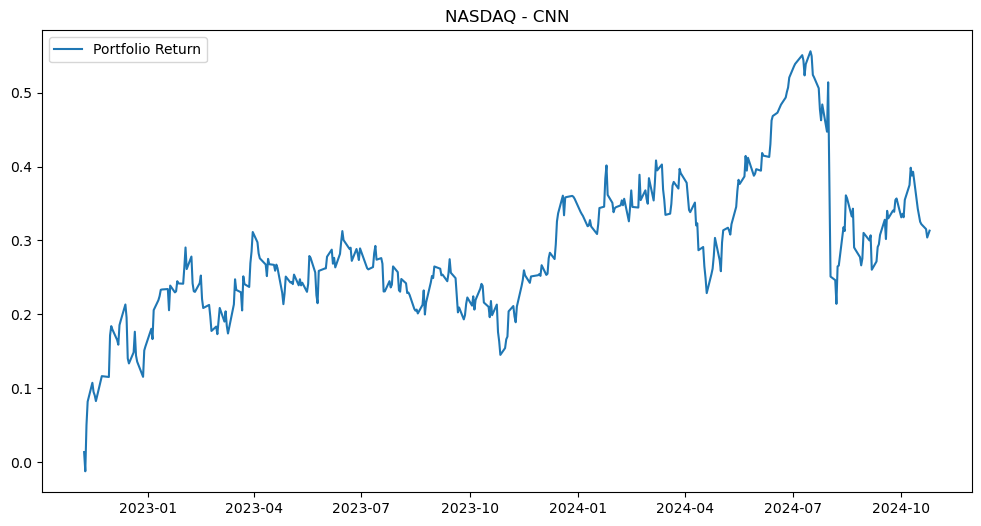}
        \caption{}
    \end{subfigure}
    \hfill
    \begin{subfigure}{0.5\textwidth}
        \includegraphics[width=\linewidth]{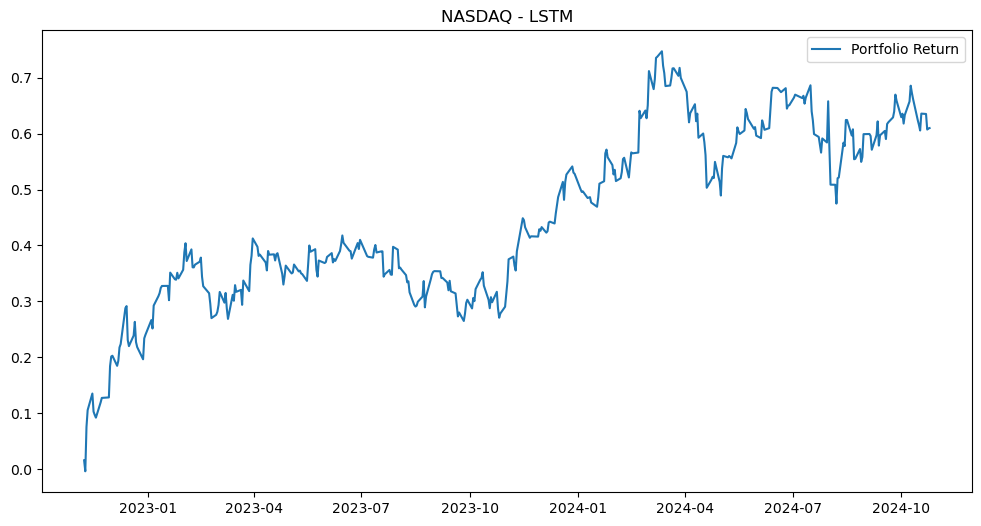}
        \caption{}
    \end{subfigure}
    \hfill
    \begin{subfigure}{0.5\textwidth}
        \includegraphics[width=\linewidth]{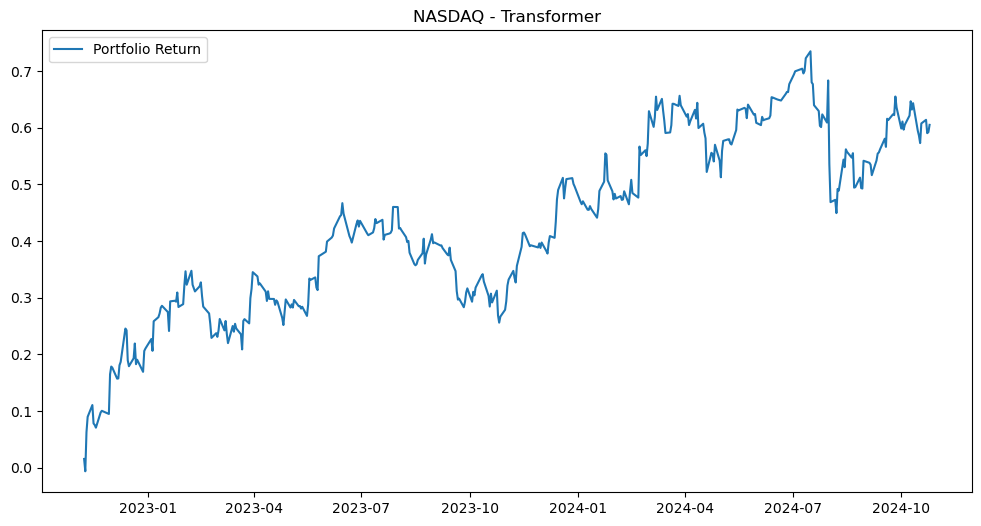}
        \caption{}
    \end{subfigure}
    \hfill
    \caption{Total return of NASDAQ stocks}
    \label{fig:multi_figs}
\end{figure}

From observation, it is noticed that the returns of stocks in both groups increased steadily over time though with some fluctuations at certain period. This indicates that the deep learning models have stable performance in predicting the returns. 

Table 2 and 3 provide the detailed performance data including total return, Sharpe ratio and maxmium drawdown.

On the S\&P 500, the Transformer model consistently outperformed the other algorithms across all metrics. Its highest total return and Sharpe Ratio, coupled with the lowest maximum drawdown, demonstrate its effectiveness in capturing market trends and managing risk in this index.

In contrast, on the NASDAQ, the LSTM model slightly outperformed the Transformer in total return and Sharpe Ratio, and also had the lowest maximum drawdown. This suggests that the LSTM's capability to model temporal dependencies was particularly effective for the NASDAQ's market dynamics during the evaluation period.

The CNN and MLP models generally underperformed compared to the Transformer and LSTM on both indices, with lower returns and higher drawdowns. The CNN's performance was notably lower on the NASDAQ, indicating that its feature extraction capabilities were less effective for this index.

\begin{table}[h!]
    \centering
    \begin{tabular}{|l|c|c|c|}
        \hline
        \textbf{Algorithm} & \textbf{Total Return} & \textbf{Sharpe Ratio} & \textbf{Max Drawdown} \\
        \hline
        MLP          & 0.514871     & 1.598638     & 0.200864     \\
        CNN          & 0.762420     & 1.955090     & 0.168542     \\
        LSTM         & 0.690007     & 1.914409     & 0.150398     \\
        Transformer  & 0.792059     & 2.270424     & 0.137142     \\
        \hline
    \end{tabular}
    \caption{Portfolio of S\&P500 Stocks Performance}
    \label{tab:performance}
\end{table}

\begin{table}[h!]
    \centering
    \begin{tabular}{|l|c|c|c|}
        \hline
        \textbf{Algorithm} & \textbf{Total Return} & \textbf{Sharpe Ratio} & \textbf{Max Drawdown} \\
        \hline
        MLP          & 0.496401     & 1.251650     & 0.223479     \\
        CNN          & 0.313203     & 0.805566     & 0.219682     \\
        LSTM         & 0.609962     & 1.351239     & 0.155910     \\
        Transformer  & 0.605086     & 1.305926     & 0.164557     \\
        \hline
    \end{tabular}
    \caption{Portfolio of NASDAQ Stocks Performance}
    \label{tab:nasdaq_performance}
\end{table}

\section{Conclusion}

This empirical study has explored the efficacy of various deep learning architectures—namely MLP, CNN, LSTM, and Transformer—in the domain of stock portfolio allocation. By meticulously selecting and analyzing a balanced sample of stocks from both the S\&P 500 and NASDAQ indices over a decade-long period, the research provides a comprehensive assessment of deep learning's potential in enhancing portfolio performance.

Our findings indicate that deep learning models, particularly LSTM and Transformer architectures, demonstrate a superior capability in predicting daily stock returns compared to traditional models. These predictions, when integrated into a portfolio construction framework employing Mean-Variance Optimization, resulted in portfolios that not only achieved higher returns but also exhibited improved Sharpe ratios and reduced maximum drawdowns. Specifically, portfolios derived from NASDAQ stocks showed slightly higher volatility but benefited from the robust predictive power of deep learning models, leading to enhanced risk-adjusted returns. Conversely, S\&P 500-based portfolios benefited from the stability and lower volatility inherent in large-cap stocks, further amplified by the sophisticated return predictions from deep learning algorithms.

The comparative analysis between the two stock groups underscores the versatility of deep learning approaches across different market segments. The dynamic adjustment of portfolios based on predicted returns proved effective in adapting to market fluctuations, thereby safeguarding against significant losses and capitalizing on positive trends. The visualization of portfolio returns during the test period corroborates these quantitative metrics, illustrating consistent performance improvements facilitated by deep learning-driven allocation strategies.

However, the study is not without limitations. The reliance on historical data from Yahoo Finance, while extensive, may not capture all market nuances, and the random selection of stocks, though methodologically sound, could introduce sample bias. Additionally, the computational complexity of deep learning models poses scalability challenges for real-time portfolio management in larger stock universes.

Future research could address these limitations by incorporating a broader array of features, including macroeconomic indicators and sentiment analysis, to further refine return predictions. Moreover, exploring ensemble methods that combine multiple deep learning architectures could enhance predictive accuracy and portfolio robustness. Investigating the application of reinforcement learning for portfolio optimization may also offer promising avenues for achieving dynamic and adaptive investment strategies.

In conclusion, this study affirms the significant role that deep learning can play in stock portfolio allocation, offering tangible improvements in returns, risk management, and overall portfolio performance. As financial markets continue to evolve, the integration of advanced machine learning techniques will undoubtedly become increasingly pivotal in shaping sophisticated investment strategies.

\end{document}